\documentclass[11pt]{article}
\usepackage[a4paper,margin=30mm]{geometry}
\usepackage[T1]{fontenc}
\usepackage{lmodern,microtype,amsmath,amssymb,booktabs,tabularx,float,xcolor,enumitem,xurl,setspace}
\usepackage{graphicx,array}
\usepackage[numbers,square,sort&compress]{natbib}
\usepackage[hidelinks]{hyperref}
\hypersetup{pdfauthor={Yi Lu, Qianshi Lin, Yumeng Yang, Hongxia Wang},pdftitle={Measuring and Charging for Retained Context in Long-Context LLM Services}}
\newcommand{\ind}{\mathbb{I}}
\begin{document}
\begin{center}
{\Large\bfseries Measuring and Charging for Retained Context in Long-Context LLM Services\par}
\vspace{0.6em}
\vspace{0.8em}
Yi Lu\textsuperscript{*}, Qianshi Lin, Yumeng Yang, Hongxia Wang\par
\vspace{0.5em}
{\small School of Statistics and Data Science, Nanjing Audit University,\\
Nanjing 211815, Jiangsu, China\par}
\vspace{0.4em}
{\small \textsuperscript{*}Corresponding author: \href{mailto:luyi06@gmail.com}{\nolinkurl{luyi06@gmail.com}}\par}
\end{center}
\section*{Abstract}
Long-context large language model (LLM) services charge for repeated context, yet billing weights need not reflect its contribution to answer quality. We present a three-layer framework separating content relevance, observable proxies and auditable positional weights. Context-ablation experiments show that relevance dominates position and that distance-decay families do not outperform a relevance-only baseline. A measured serving stack shows that cache hits reduce repeated-prefix compute cost by over two orders of magnitude at 65,536 tokens, while cold resends constrain sustainable discounts. Cross-provider measurements separate cache value from tokenisation and list-price differences. Calibrated simulations reduce the 99th percentile of session charges by 31--42\%, although revenue-matched nonlinear tariffs achieve similar reductions and median charges increase. The framework provides a bounded per-token accounting rule and measured deployment constraints, rather than a validated utility estimator.

\noindent\textbf{Keywords:} large language models; cloud service pricing; context caching; inference cost; service measurement; auditable billing
\clearpage
\section{Introduction}
The rapid expansion of LLM context windows has enabled document-level reasoning, repository-scale coding assistance and long-running agent workflows. In a multi-turn application, a request contains new tokens together with a large prefix of previously observed context, so commercial APIs distinguish uncached input, cached input, output and sometimes internal reasoning tokens. Caching lowers the price of reusing an identical prefix, but the billing weight assigned to historical context remains only weakly connected to its contribution to the quality of the current answer.

This motivates a measurement and accounting question for cloud-based inference services: should every retained unit of context carry the same billing weight, or should the price reflect its expected contribution? Distance alone is not a complete measure of relevance---a remote contractual clause may remain decisive while a recent aside may carry little value---so distance-based decay must be treated as a testable approximation rather than as a property implied by Transformer attention.

We develop a weighting framework for historical input in which the weight has three layers: an \emph{ideal} layer, a function of the block's measured contribution to the task; a \emph{proxy} layer, a function of observable content features; and an \emph{auditable} layer, a function of position alone that a billing system can compute without inspecting content. The deployable instance analysed here is the auditable layer, the ablation study of Section~\ref{sec:evidence} measures the ideal layer, and the proxy layer is left to future work. The framework makes three distinctions: the lifetime charge of one historical token can be bounded; the historical component of a single request can be bounded under stable arrivals; and total session expenditure normally grows without converging to a ceiling.

These distinctions frame three questions. \textbf{RQ1}: does the contribution of a historical block decline systematically with its distance from the request, and if so how? \textbf{RQ2}: among candidate decay families, which best predicts held-out quality across tasks and models? \textbf{RQ3}: under what conditions on serving cost and demand response does a context-weighted tariff improve provider profit and stakeholder welfare relative to uncached and cache-based linear pricing?

Three measurements and a calibrated simulation carry the argument: a context-ablation study over legal contracts, code and agent transcripts addresses RQ1 and RQ2, while a calibrated simulation built on the measured price schedules, a measured serving stack and a two-provider switching experiment addresses RQ3. The ablation answers RQ1 negatively as posed: contribution varies with position but not monotonically, and no distance-decay family beats a relevance-only specification. The simulation locates the demand response the provider condition requires, and the two measurements establish where a discounted historical charge is sustainable and how much of the switching premium a portable rule recovers. One framing point follows: the instance analysed is positional while the measured ideal is relevance-based, so the gap is a deployment constraint rather than a result. The mechanism's robust benefit is a smaller upper tail rather than a lower average bill, since the median calibrated session pays more. For cloud-service management, these results connect an auditable charging rule to measured context quality, cache-dependent serving cost and cross-provider transfer charges. Deployment across providers additionally requires verifiable block histories.

\section{Related Work}
Self-attention makes long sequences expensive \citep{vaswani2017}; long-context inference research improves execution and cache management \citep{dao2023,kwon2023} or reduces the information processed and retained \citep{child2019,acharya2025,lin2025}, and none of this implies that attention weights are monetary values, or that semantic utility must decline geometrically with distance. Long-context applications span autonomous agents, finance, health care, education, software engineering and legal research \citep{wang2024,xi2025,boiko2023,li2023,thirunavukarasu2023,kasneci2023,kazemitabaar2023,katz2024}, and their relevance profiles differ sharply---an argument against reading a single provider-chosen discount rate as a measure of model intelligence. Our contribution is narrower: it isolates the context that reappears across requests and specifies the measurements needed to connect its price to quality and cost. The closest neighbour is Bergemann, Bonatti and Smolin \citep{bergemann2025}, who characterise a monopolist's menu of contracts over token allocation, fine-tuning and usage and show that two-part tariffs implement the optimum: they take the mapping from tokens to quality as given and ask what to charge for it, whereas this paper takes the incumbent's charge as given and measures that mapping for tokens retained from earlier requests rather than newly allocated. The age of a retained block is a screening dimension their menu does not carry, and the cost and demand evidence collected here is the kind of input their optimality conditions require. Extended positioning---nonlinear and two-part tariffs \citep{mussa1978,maskin1984,wilson1993,oi1971,rochet2002,armstrong1999,rochet2006}, information goods \citep{shapiro1999,bakos1999,sundararajan2004}, cloud and software pricing \citep{choudhary2007,armbrust2010,buyya2009,weinhardt2009}, platform governance \citep{tiwana2010,boudreau2010,hagiu2015,parker2016} and the economics of LLM pricing \citep{bergemann2025}---is in Appendix~F.

\section{Pricing Framework and Bounds}
Let $k=1,\ldots,n$ index rounds. Newly introduced input, visible output retained in future context and non-persistent reasoning tokens are $I_k$, $O_k$ and $R_k$; define $\Delta_k=I_k+O_k$ and $H_k=\sum_{j\le k}\Delta_j$, with $H_0=0$. Let $W$ be the maximum historical tokens admitted to a request and $P_{\rm in}$, $P_{\rm cache}$, $P_{\rm out}$, $P_{\rm reason}$ the category-specific per-token prices. For a repeated prefix, the cache benchmark is
\begin{equation}
C_k^{\rm cache}=P_{\rm in}I_k+P_{\rm cache}L_k+P_{\rm out}O_k+P_{\rm reason}R_k,
\label{eq:cache}
\end{equation}
where $L_k=\min\{H_{k-1},\max(0,W-I_k)\}$; if only a fraction $h_k$ of the history is cached, its charge is $[h_kP_{\rm cache}+(1-h_k)P_{\rm in}]L_k$, and $h_k$ should come from usage records rather than from a fixed assumption.

The proposed charge replaces the cache rate on history by a weight $w\in[0,1]$ applied to the chargeable overlap $A_{jk}$ of block $j$ with the admissible window:
\begin{equation}
\begin{aligned}
C_k^{w}={}&P_{\rm in}I_k+P_h\sum_{j=1}^{k-1}A_{jk}w(D_{jk},x_{jk};\theta)\\
&+P_{\rm out}O_k+P_{\rm reason}R_k,
\end{aligned}
\label{eq:general}
\end{equation}
where $P_h$ is the historical base price, $D_{jk}$ the token distance of block $j$ from the request, $x_{jk}$ task characteristics and $A_{jk}$ the interval overlap with the window boundary (stated in closed form in Appendix~B). Candidate weights include exponential, power-law and hyperbolic forms, with the turn-based geometric weight $w(a)=r^a$, $a=k-1-j$, the special case used in Section~\ref{sec:evidence}.

Three bounds hold for the geometric case and carry over to any $w\le1$. The lifetime charge of one token over $n+1$ appearances is
\begin{equation}
B_n=P_h\sum_{a=0}^{n}r^a=P_h\frac{1-r^{n+1}}{1-r}\;\longrightarrow\;\frac{P_h}{1-r},
\label{eq:lifetime}
\end{equation}
so the repeated charge attributed to a single token converges to a finite limit. Under constant arrivals $\Delta$ the historical part of the round-$k$ charge is
\begin{equation}
C_{k,\rm hist}^{r}=P_h\Delta\sum_{a=0}^{k-2}r^a=P_h\Delta\frac{1-r^{k-1}}{1-r},
\label{eq:marginal}
\end{equation}
rising monotonically to $P_h\Delta/(1-r)$, and cumulative historical expenditure is
\begin{equation}
\sum_{k=1}^{N}C_{k,\rm hist}^{r}=\frac{P_h\Delta}{1-r}\left[N-\frac{1-r^N}{1-r}\right],
\label{eq:cumulative}
\end{equation}
which is asymptotically linear in $N$: the session total is bounded neither by the token-level nor by the request-level bound. Cost coverage is a separate requirement, $C_k^{w}\ge G_k$ for provider variable cost $G_k$; with demand $D(P_h,\theta)$ profit is $\Pi=D\,[\mathbb E(C_k^{w})-\mathbb E(G_k)]$, so a claim that discounting raises profit needs evidence on both serving cost and demand, and non-negative weights alone do not guarantee that the condition holds.

\subsection{The tariff as a price schedule}
Written as a price schedule, the proposal sets the price of a retained token by its age. For a block of age $a$ the average price per token is
\begin{equation}
\alpha(a)=P_h\,w(a),
\label{eq:avgprice}
\end{equation}
and because the charge is linear in tokens within a block, $\alpha(a)$ is also the marginal price of retaining one more token of that age. The geometric instance is $\alpha(a)=P_h r^a$, a schedule declining geometrically in age. The three tariffs compared in Section~\ref{sec:sim} are three coordinates on the same object: the cache benchmark is the flat case $\alpha=hP_{\rm cache}+(1-h)P_{\rm in}$; a two-part tariff keeps $\alpha$ flat and adds a fee $F$ orthogonal to the usage margin; a length tier makes $\alpha$ a step function of total retained volume rather than of age. Two properties follow from the shape of $\alpha$. It is non-increasing in $a$, so the schedule is a quantity discount along the age coordinate---not the coordinate along which the classical results are stated, since Oi's two-part tariff \citep{oi1971} and Wilson's quantity discounts \citep{wilson1993} make the price depend on how much is bought, whereas here it depends on how long it has been held. And $\alpha(a)\le P_h\le P_{\rm in}$ whenever the historical base price does not exceed the uncached rate---the calibration sets the two equal, so the inequality is tight there---which is why the weighted tariff never prices history above the uncached rate while the incumbent cache prices it at $P_{\rm cache}\ll P_{\rm in}$. The schedule above is written on the age coordinate; the general form of Equation~\eqref{eq:general} weights a block by its token distance $D_{jk}$ instead, and the two coincide only when every block of a round carries the same age, as they do for a dialogue but not for a document delivered in one round. Section~\ref{sec:sim} prices the age form and Section~\ref{sec:ctx} tests the distance form. Which schedule is cheaper for a given history therefore depends on its age, and the crossover in Section~\ref{sec:sim} is where the two meet.

\section{When the Tariff Works}
\label{sec:mechanism}
The results here are accounting identities and feasibility conditions rather than equilibrium theorems: they state what must hold for a weighted tariff to be well defined, bounded and cost-covering, and which quantities an empirical study must supply. The proofs and the formal statements of those conditions are in Appendix~B.

For any finite window and bounded per-round arrivals the charge per request satisfies
\begin{equation}
C_k^w\le P_{\rm in}\overline I+P_hW+P_{\rm out}\overline O+P_{\rm reason}\overline R,
\label{eq:requestbound}
\end{equation}
and if arrivals and the retained composition become stationary the charge converges to a stationary distribution while cumulative expenditure grows at the finite long-run rate $\mathbb E[C_k^w]$ (Proposition~1). Exceeding the window therefore creates no extra charge for evicted tokens; it changes the composition of the charged state (Corollary~1). Compression is preferred to eviction exactly when the value a summary preserves exceeds its one-time cost plus the discounted future cost of retaining it (Proposition~2); a long window changes when that choice is feasible but does not remove the trade-off.

For the same history, portable weighting beats cold linear input at a destination if and only if
\begin{equation}
P_{h,m}\sum_{j<k}A_{jk}^{(m)}w_{jk}+\kappa<P_{{\rm in},m}\sum_{j<k}A_{jk}^{(m)},
\label{eq:switchcondition}
\end{equation}
equivalently $\bar w_A<P_{{\rm in},m}/P_{h,m}-\kappa/(P_{h,m}L_{m,k})$ for the token-weighted mean weight $\bar w_A$ (Proposition~3): long histories amortise the fixed migration cost $\kappa$ over more retained tokens; the relevant comparison is the destination's cold-start charge rather than the incumbent's warm cache, and Section~\ref{sec:switching-measured} measures both sides of that comparison. The migration cost $\kappa$ is a design assumption rather than a measurement: it is set to $1.25\%$ of the cold-start charge for the transferred history, a relative magnitude carried over from the original placeholder calibration; the paper does not measure it. If per-block marginal serving cost were observable, a usage price equal to marginal cost would implement the efficient retention choice and a fixed access fee could recover fixed cost without distorting it (Proposition~4). The assumption is strong: in production serving cost is shared across concurrent requests, so the result is an idealised benchmark rather than an implementable rule. The proposed regime weakly improves on cache pricing for provider, application and end user together if the corresponding three conditions hold with at least one strict (Proposition~5, which records a decomposition rather than proving a substantive implication); lower historical prices alone are insufficient, since quality degradation can violate the user condition, weak demand response the provider condition, and a higher median or migration cost the application condition. Of the three, the user condition is the one this study does not evaluate: the provider condition is quantified by the demand elasticity it requires and the application condition by the cost distribution, but the user condition depends on a quality response to the tariff that no experiment here measures, and it is also the condition a quality loss could break. This proposition turns the multi-party benefit claim into three separate empirical tests: it explains why the paper reports both mean and upper-tail application costs, and why model quality must be held constant in price comparisons.

For the provider, the condition has a directly testable form---a usage-growth threshold---
\begin{equation}
\frac{D_1}{D_0}>\frac{p_0-g_0}{p_1-g_1},
\label{eq:threshold}
\end{equation}
and Section~\ref{sec:evidence} shows the demand response it requires; it avoids assuming that every price reduction is profitable. For an application developer, Equation~\eqref{eq:marginal} implies that the historical component of each request approaches a constant under stable arrivals, so the benefit is a stabilised per-round cost and a thinner tail rather than an identical price for every conversation. For a user, a portable rule would charge transferred history by verifiable age or contribution weight rather than by local cache state, lowering the switching charge and increasing contestability---but a destination must be able to verify blocks, their age and the applicable weight, so the proposal is a pricing-plus-standardisation mechanism rather than a feature current APIs support. Table~S6 states the three conditions and the measure each outcome is assessed on; each is stated relative to current cache pricing, which is the baseline regime throughout.

\noindent One consequence of the general tariff is substantive rather than formal: because the tariff separates persistent state from ephemeral computation, an agent workload requires dialogue, tool output and summaries to be weighted separately, with reasoning tokens charged only in the current round. That modal separation drives the high premium on persistent tool state reported in Section~\ref{sec:evidence}, where charging tool output at the historical base price is exactly the case in which the weighted tariff loses to a strong incumbent cache. The corresponding formulations are in Appendix~A.

\section{Measurements and Simulation}
\subsection*{Use of generative artificial intelligence}
ChatGPT (OpenAI) was used to assist with language polishing and with developing and revising data-processing scripts for data cleaning and aggregation. The authors take responsibility for the data-processing procedures and for the accuracy and integrity of the manuscript. The use of large language models as experimental systems is described separately in the relevant experimental subsections.

\label{sec:evidence}
\label{sec:measured}
Three measurements and one calibrated simulation supply what the framework needs, in the order in which they constrain it: what history contributes to quality (Section~\ref{sec:ctx}), what serving costs (Section~\ref{sec:cost}), what switching costs (Section~\ref{sec:switch}), and what the tariff is worth once those three enter it as inputs (Section~\ref{sec:sim}). The first three are measurements on live or local systems, the last a calibrated simulation, each labelled as such throughout.

\subsection{What history contributes}
\label{sec:ctx}
We ran a context-ablation study on 100 CUAD legal contracts (median $9{,}600$ tokens, maximum $63{,}200$), supplemented by code-completion tasks and 100 synthetic agent transcripts. Each document is evaluated under six conditions---full context; removal of the answer block; that block moved nearer to or farther from the question; and replacement by an equal-length summary or distractor---with position randomised independently of relevance, using \texttt{deepseek-flash} at temperature zero ($5{,}110$, $830$ and $600$ calls, respectively, in the main ablation, the window-policy study and the agent runs).

\begin{figure}[H]
\centering
\includegraphics[width=\textwidth]{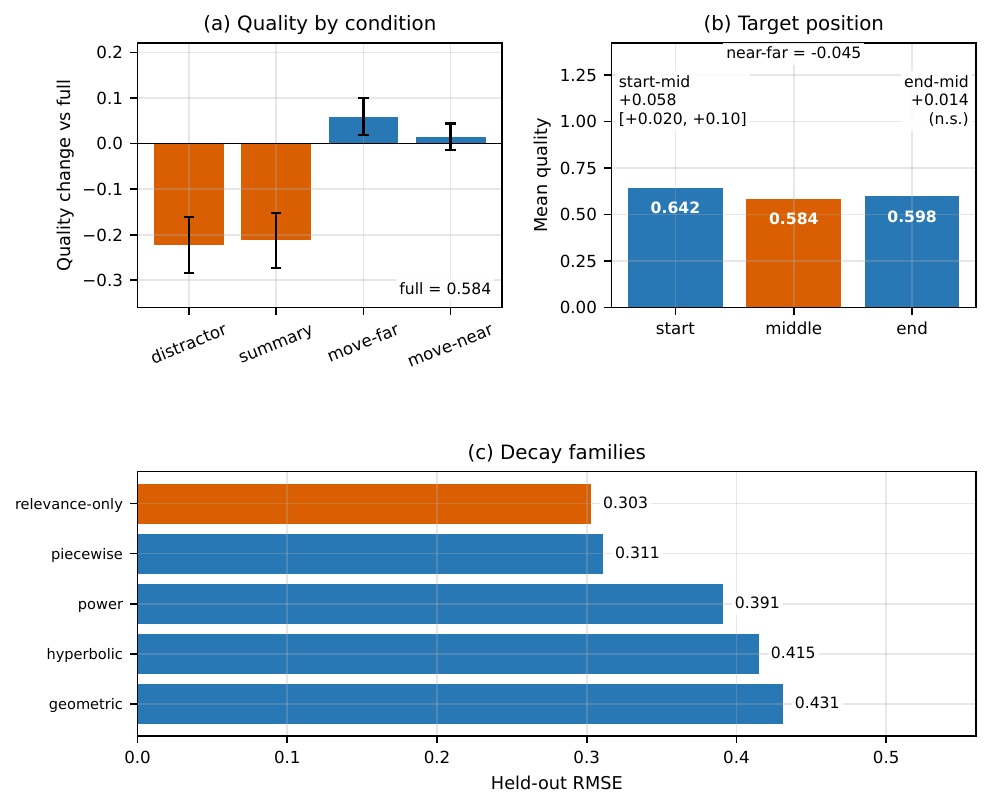}
\caption{Context utility on CUAD contracts \textbf{[measured]}, $n=100$ per condition. \emph{(a)} Quality change against full context, with the 95\% paired-contrast interval of each ablation. \emph{(b)} Mean quality by target position, with both contrasts against full. \emph{(c)} Held-out test RMSE by decay specification: the relevance-only baseline is lowest, the geometric family highest.}
\label{fig:ctx}
\end{figure}

Content dominates. Replacing the answer block by a length-matched distractor or by a summary lowers the quality score by $0.222$ (95\% CI $[-0.284,-0.162]$) and $0.211$ ($[-0.274,-0.153]$) respectively, roughly four times the largest positional contrast ($+0.058$) and five times the near--far gap (Figure~\ref{fig:ctx}, panel a). Position matters but is not monotone, consistent with reported position effects \citep{liu2024}: quality is highest at the beginning and end of a document and lowest in the middle (panel b). Most consequential for the tariff, no distance-decay family predicts held-out quality better than a baseline containing only a content-relevance term---the relevance-only specification reaches a test RMSE of $0.303$, against $0.311$ for a piecewise-linear family, $0.391$ for power-law, $0.415$ for hyperbolic and $0.431$ for exponential (geometric) decay (panel c). Read against the relevance-only baseline, the three monotone families are significantly worse on the held-out test set (clustered bootstrap intervals on the RMSE difference: $+0.089$ $[+0.037,+0.141]$ for power-law, $+0.112$ $[+0.057,+0.168]$ for hyperbolic and $+0.128$ $[+0.072,+0.185]$ for exponential), whereas the piecewise-linear family is not ($+0.009$ $[-0.004,+0.023]$)---and a replication on a second model family of 50 contracts reproduces both results. That baseline is deliberately weak: it predicts a two-level mean over a binary label rather than a graded gradient. The comparison therefore shows that distance adds no predictive power over that mean, not that relevance explains the residual variance. Two further qualifications matter here. The five specifications are compared on the long-document slice, which is the component with the most weight; over all tasks in the same study a piecewise-linear distance family is selected, though its held-out advantage over the relevance-only baseline is not significant ($+0.009$, $[-0.004,+0.023]$). And the design has limited power on distance itself: the clean contrast, which keeps the answer block in place and displaces only the target, leaves $n=1$--$4$ observations beyond $20{,}000$ tokens, so what the test rules out is a monotone family that beats the content-relevance baseline, not a weak non-monotone distance signal. The implication for $w$ is direct: a monotone geometric discount is not the form that best tracks measured contribution, so the general specification retains information that the geometric special case discards. What the evidence supports is a weight anchored to content relevance, with distance entering only as a weak and non-monotone control. Because relevance is not observable to a billing system while position is, a deployed tariff must trade accuracy against auditable simplicity.

Two further blocks of the same study bear directly on the tariff. Under window management, compressing irrelevant content while the answer block is retained leaves quality essentially unchanged---$0.497$, $0.497$, $0.484$ and $0.487$ at no compression and at $50\%$, $75\%$ and $90\%$ retention, a spread of $0.013$---so the compression the tariff has to pay for is close to quality-neutral on this task. In the agent transcripts, by contrast, every early observation is irreplaceable: success falls from $1.00$ to $0.00$--$0.01$ when any single early observation is deleted, so its contribution is about one and its age carries no information at all.

One consequence is worth pricing rather than asserting. Because the answer-bearing block sits at a known position, the gap between the two layers can be measured on the charge side even though its quality consequence cannot. Treating a contract as history delivered in a single round---so that every block shares one age and the turn-based weight gives them all the same factor---the geometric weight levies a median $18.8$ times the charge a relevance-anchored weight would levy with irrelevant blocks at zero weight ($5.3$--$48.1$ across the $100$ contracts, tenth to ninetieth percentile), and $6.5$ times when those blocks carry a weight of $0.10$, which is itself $16\%$ above the cache benchmark. Both ratios are upper bounds on the gap, because the relevance-anchored side is an oracle: it charges in full only the block that actually carries the answer, which a billing system cannot identify. Read against the framework, this says more than that the approximation over-charges. A tariff has to recover the cost of holding and serving history, and a zero-floor relevance anchor collects $40\%$ of the cache benchmark, roughly half of what the weighted tariff collects, which scales the coverage ceiling of Section~\ref{sec:cost} down by the same factor. Cost recovery, not efficiency, is therefore the first reason history is charged at all, and it is why the deployable weight puts a floor under blocks it cannot identify as relevant: relevance anchoring is not only harder to audit, it collects too little to be assumed cost-covering at the longest windows.

\subsection{What serving costs}
\label{sec:cost}
We measured the variable cost of serving a request on an open-source stack: Qwen2.5-1.5B-Instruct at four-bit precision on one consumer GPU (RTX 3050 Laptop, 4\,GB) through \texttt{llama.cpp}, at context lengths of $4{,}096$, $16{,}384$, $32{,}768$ and $65{,}536$ tokens (written 4K, 16K, 32K and 64K below), with thirty timed repetitions per cell. Costs combine \$0.05 per GPU-hour, \$0.00875 per hour of depreciation and \$0.077 per kWh of electricity.

\begin{figure}[H]
\centering
\includegraphics[width=\textwidth]{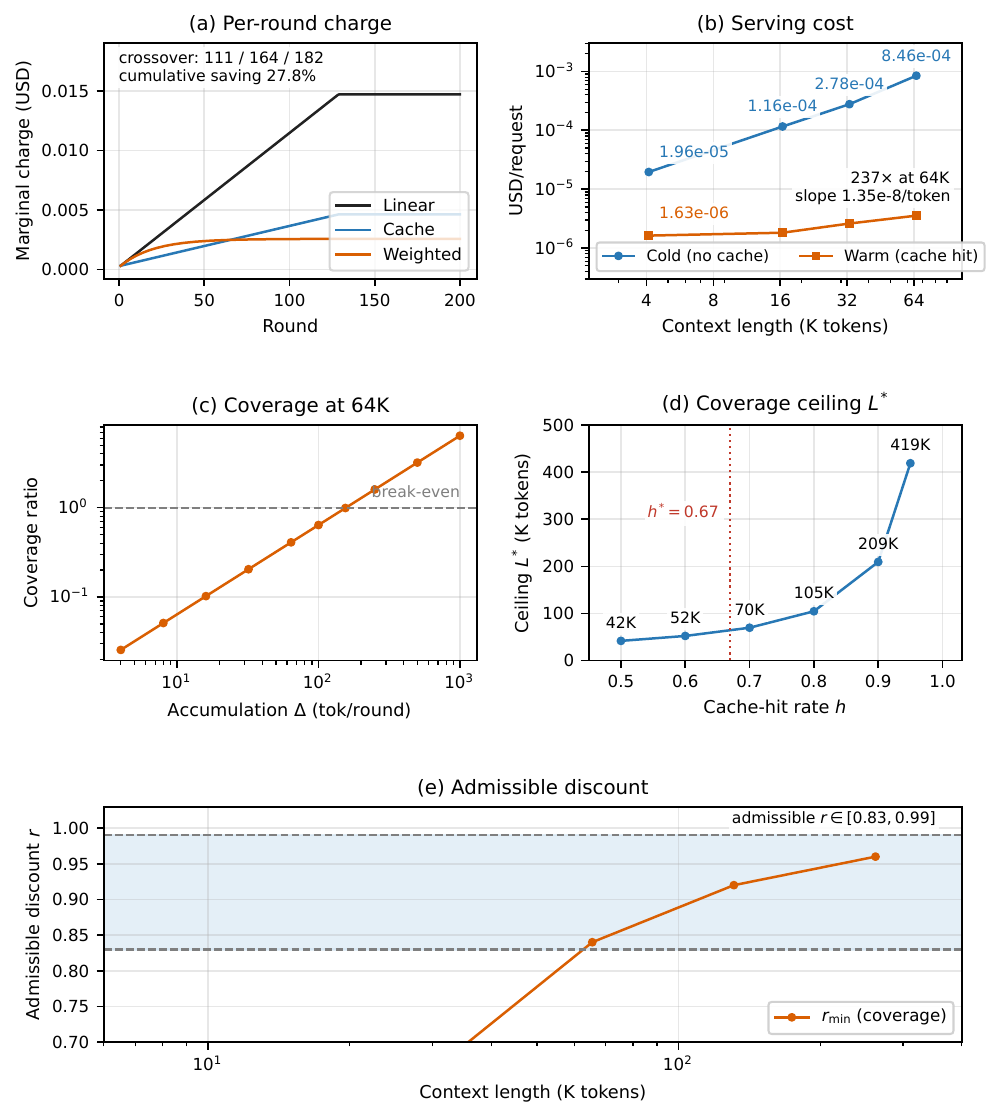}
\caption{Costs and coverage. \emph{(a)} Per-round charge under the three tariffs \textbf{[calibrated]}. \emph{(b)} Serving cost per request against context length, cold and warm \textbf{[measured]}---warm cost runs from $\$1.63\times10^{-6}$ at 4K to $\$3.56\times10^{-6}$ at 64K; the four cold values are labelled. \emph{(c)} Coverage ratio of the historical charge against the expected cold re-send cost, against accumulation \textbf{[derived from the calibration and the measured slope]}. \emph{(d)} Historical-component coverage ceiling against the cache-hit rate ($\Delta=250$, $r=0.90$) \textbf{[derived]}. \emph{(e)} Admissible discount against context length ($\Delta=250$, $h=0.80$); the band is the interval implied by coverage and competition \textbf{[derived]}.}
\label{fig:cost}
\end{figure}

Two structural facts emerge, and both concern the shape of serving cost rather than its absolute level (Figure~\ref{fig:cost}, panels b--e). A cache hit collapses the cost of a repeated prefix: at 64K the median cold request costs $\$8.46\times10^{-4}$ and takes 48.1\,s, while the warm request costs $\$3.56\times10^{-6}$ and takes 0.21\,s---$237\times$ in cost and $226\times$ in latency. Warm cost is therefore almost independent of context length, whereas cold cost grows near-linearly, with a measured slope of $\$1.35\times10^{-8}$ per token over the full 4K--64K range; the 64K endpoint is mildly super-linear, plausibly from memory pressure on a 4\,GB card, and dropping it lowers the slope to $\$9.0\times10^{-9}$, which would move the coverage ceiling quoted above outward by about $50\%$, not inward. A third fact concerns load: throughput rises with concurrency across the three levels tested, from $11.8$ to $26.7$ to $53.8$ requests per second at concurrency one, two and four respectively, so peak load can be absorbed at low concurrency without a proportional increase in cost.

Substituting these measurements into the coverage condition of Section~\ref{sec:mechanism} separates two regimes: in the steady state the historical charge is covered with a wide margin, while the binding case is that of a cold re-send, where the historical charge is bounded by $P_h\Delta/(1-r)$ but the prefill cost grows linearly. A coverage ceiling therefore exists, of order $2$--$4\times10^{4}$ tokens if every round were a cold re-send and rising to roughly $10^{5}$ at a hit rate of $0.8$ and $r=0.90$, so a 64K context requires $h$ above roughly $0.67$. The hit rate is an input to this calculation rather than a measurement, and the three values that appear in this paper are not interchangeable: the calibrated baseline assumes $h=0.80$ for mixed traffic, the switching experiment measures $h\approx0.999$ for an adjacent repeated prefix, and the ablation study's own runs had a $56\%$ cache-hit rate. In the steady state the margin is large at every hit rate the measurements support, so the constraint binds only in the cold re-send, where $h$ is zero by construction. The same constraints bound the discount itself: coverage requires $r\ge0.83$ at 64K under the baseline accumulation, while competition requires $r\le0.99$, giving an admissible interval of $[0.83,0.99]$ (panel e). That bound is computed on the historical component of the charge alone, which is the component $r$ can affect; under the whole-round reading of Section~\ref{sec:mechanism}, in which the same round's uncached new input and output are counted in the numerator, the bound vanishes at this point, because that revenue alone exceeds the expected cold re-send. The measurement covers compute only---the key--value cache of a 64{,}000-token context is about $1.9$~GB in half precision, and that holding cost grows with length, so including it would tighten rather than relax the ceiling---and, because the stack is a small model on a consumer GPU, these results establish scaling relationships rather than the absolute cost of a frontier deployment. Its generations were also shorter than the lengths a production service would request, so the output component of the measured cost is a lower bound.

\subsection{What switching costs}
\label{sec:switch}
\label{sec:switching-measured}
We sent an identical history to an incumbent API with a warm cache (\texttt{deepseek-flash}) and a destination API with a cold cache (\texttt{qwen3.8-flash}) at $10{,}000$, $100{,}000$ and $200{,}000$ tokens, twenty rounds per length. Each round comprises a first cold send to each provider followed by repeated warm sends, so the 120 cold and 240 warm requests reported below come from sixty rounds in total. A round-unique marker was embedded in every sentence to keep the destination cold: both providers match cached prefixes in $1{,}024$-token blocks, so a marker at the start alone would still have hit on deeper repeated spans. All 120 cold requests recorded zero cached tokens and all 240 warm requests positive ones.

\begin{figure}[H]
\centering
\includegraphics[width=\textwidth]{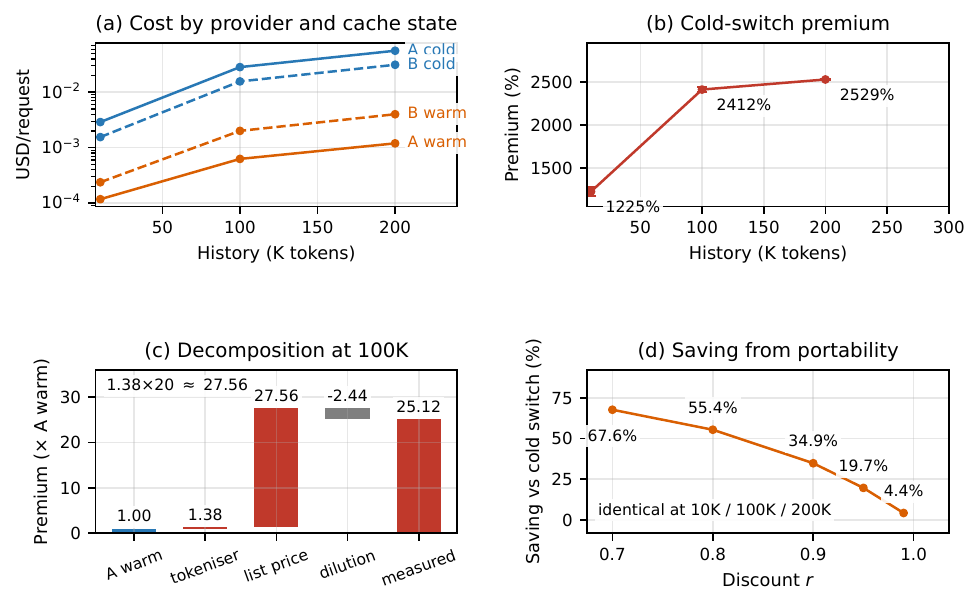}
\caption{Cross-provider switching \textbf{[measured]}. \emph{(a)} Cost per request by provider and cache state. \emph{(b)} Cold-switch premium, 95\% intervals clustered by round. \emph{(c)} Decomposition at $100{,}000$ tokens: tokeniser ratio $\times$ price ratio, diluted by the incumbent's length-independent warm floor. \emph{(d)} Savings from portability relative to the destination's cold charge, identical at all three lengths.}
\label{fig:switch}
\end{figure}

The measured cold-switch premium $(C_{B,\rm cold}-C_{A,\rm warm})/C_{A,\rm warm}$---the destination's cold charge against the incumbent's \emph{warm} cache---is $1225\%$ (95\% CI $[1176,1280]$) at $10{,}000$ tokens, $2412\%$ ($[2382,2438]$) at $100{,}000$ and $2529\%$ ($[2521,2537]$) at $200{,}000$ (Figure~\ref{fig:switch}, panels a--b). The premium rises with length because the incumbent's warm charge retains a length-independent floor: the new question and the generated answer are billed at the uncached rate every round. The premium decomposes into three factors (panel c): at $100{,}000$ tokens the destination tokenises the same text into $1.38\times$ as many tokens, its uncached input price is about $20\times$ the incumbent's peak cached-input rate, and the two factors combined, about $28\times$, are diluted to the measured value by that floor. The premium is therefore the value of the incumbent's cache, not a penalty on the destination. Netting the cross-provider differences out leaves no switching friction in price to explain: against the incumbent's own \emph{cold} charge the destination's is $46.8\%$, $44.3\%$ and $44.2\%$ lower at $10{,}000$, $100{,}000$ and $200{,}000$ tokens, so moving a cold history to the destination costs less than re-sending it to the incumbent. Nor is it purely a cache effect, since the text tokenises to $1.34\times$--$1.40\times$ as many tokens under the destination's tokeniser. First-token latency divides the same way and is dominated by the destination's baseline: the incumbent answers in 0.95\,s cold and 0.74\,s warm at $10{,}000$ tokens, and in 3.96\,s and 1.50\,s at $200{,}000$, whereas the destination takes 5.5\,s and 4.9\,s at $10{,}000$ and 70.6\,s and 41.4\,s at $200{,}000$.

Applying the portable-weighting rule offline to a history of ten equal blocks gives savings that are invariant to length: relative to the destination's cold charge, they are $67.6\%$, $55.4\%$, $34.9\%$, $19.7\%$ and $4.4\%$ at $r=0.70$, $0.80$, $0.90$, $0.95$ and $0.99$ respectively (panel d). The invariance is in length and not in block count: the counterfactual prices $N$ equal blocks, the saving is $1-\frac1N\sum_{a<N}r^a$, and at $r=0.95$ it is $19.7\%$ for $N=10$, $35.8\%$ at $N=20$ and $80.1\%$ at $N=100$, so the five values should be read as the saving for a history accumulated over ten rounds. The rule discounts the destination's uncached price rather than replacing its cache, so it does not beat a warm incumbent cache and is not intended to: it caps the cold-start charge when history is moved. It presumes that a destination can verify each transferred block's hash, age and token count; without such signed metadata, the charge reverts to the cold rate. The rule is a counterfactual that no provider currently offers. One collection detail bears on the amounts above: the $100{,}000$-token arm was recorded during the incumbent's idle window, when its cached-input rate is half the peak rate, and every reported amount has been recomputed on the peak schedule.

\subsection{What the tariff delivers}
\label{sec:sim}
The simulation compares three tariffs---uncached linear, prefix-cache and the weighted tariff of Equation~\eqref{eq:general}---on identical token streams. Every calibrated result is computed under all three official price schedules; Table~\ref{tab:schedules} reports them side by side and Table~S2 gives the schedules themselves. The figures of this section use the \texttt{qwen3.8-flash} schedule ($0.8$, $0.10$ and $2.7$ CNY per million tokens for uncached input, cached input and output, respectively, at $7.1$ CNY per USD). A 200-round primary session is crossed with a grid over five workload types, two windows, three hit rates and six values of $r$ (180 combinations); a 1{,}200-round memory experiment reaches the binding region at 32K, 128K and 1M (one million tokens); a Monte Carlo experiment draws 2{,}000 users. Service cost, demand response and the user distributions are calibration assumptions, not estimates from customer logs, and the full grid is in Table~S1.

The gain is not uniform across price schedules, and that is the first thing to report (Table~\ref{tab:schedules}). The 200-round session advantage runs from $3.3\%$ under the DeepSeek \textit{flash} schedule to $27.8\%$ under \texttt{qwen3.8-flash}, and the crossover from round 182 to round 111, because a deeper incumbent cache discount is harder to beat. More consequentially, the mean simulated user cost falls $11.4\%$ under \texttt{qwen3.8-flash} but \emph{rises} $10.3\%$ and $7.4\%$ under the two DeepSeek schedules, so the sign of the average effect is a property of the schedule rather than of the mechanism. What holds under all three is the tail: P99 falls $30.9\%$--$42.0\%$ and the coefficient of variation $40.5\%$--$43.5\%$. The robust claim is therefore a smaller upper tail, not a lower average bill. The magnitude of that tail gain depends on whether the application compresses at all, and on the serving stack the compression is applied to: with the proactive rule of Appendix~C switched on at $\tau=0.90$, the same $2{,}000$ users see P99 fall $32.8\%$ and the coefficient of variation $30.3\%$, against $42.0\%$ and $40.5\%$ under hard truncation, while the median rises $58.9\%$ either way (Table~\ref{tab:compressiondist}). Compression moves only the sessions long enough to fill the window, which is why it narrows the tail gain without changing its sign. On the \texttt{qwen3.8-flash} baseline---200 rounds, 250 new input and 750 retained output tokens per round, a 128K window, $h=0.80$, $r=0.95$---the weighted session costs \$0.468 against \$0.648 under the cache benchmark, a $27.8\%$ reduction, with the mean user cost down $11.4\%$, the standard deviation down $47.3\%$, the median up $58.9\%$, P99 down $42.0\%$ and the coefficient of variation down $40.5\%$ (Table~\ref{tab:developer}). The mechanism shifts cost away from extreme long sessions---a redistribution financed by a higher median. Whether that shift improves welfare depends on how session length correlates with willingness to pay, which this study does not observe.

Two standard instruments compress the same distribution, so the question is what the weighted tariff buys. The comparison has to hold revenue fixed, since the weighted tariff collects $11.4\%$ less than the cache benchmark on the $2{,}000$-user population ($27.8\%$ on the baseline session) and an instrument that collects more is not being compared on equal terms. At the weighted tariff's revenue and its P99 reduction the median rises $58.9\%$. A two-part tariff calibrated the same way needs $\alpha=0.551$ and a fixed fee of \$0.0183, and raises the median by $86.0\%$; a fixed fee is blunt because it lifts every bill rather than the long ones. A length tier is not blunt: tuned to the same revenue and the same P99, a rate charging $0.20$ times the cache rate beyond $32{,}000$ history tokens with a fee of \$0.0086 reproduces the coefficient of variation as well, so the only independent result in that comparison is the median increase, $61.5\%$ against the weighted tariff's $58.9\%$. The distribution the tariff produces is therefore not what distinguishes it; what distinguishes it is the coordinate it prices on.

A three-way comparison makes the position of the deployable instance explicit. Pricing by relevance targets contribution best---by construction, since it is defined on it---but it is not observable to a billing system, and at a zero floor it collects only $40\%$ of the cache benchmark (Section~\ref{sec:ctx}). Pricing by total volume is observable but the least targeted of the three, because it cannot separate a long document that is fresh and relevant from one that is old and irrelevant. Pricing by age is observable and lies between the two: Section~\ref{sec:ctx} finds age to be a weak and non-monotone predictor of contribution, so the age weight is a better-targeted approximation than volume and a worse-targeted one than relevance, and what it buys in return is that it is computable, verifiable and transferable.

On the provider side, with variable cost assumed at $55\%$ of baseline revenue, the weighted tariff's profit exceeds the cache-pricing baseline only once demand elasticity reaches about $3.0$---a high bar for a mature digital service, so the provider gain is conditional rather than typical.

Coverage is likewise not uniform across windows: at the calibrated hit rate of $0.8$ and the baseline $r=0.95$ the historical charge covers the expected cold re-send by $6.4$ times at 32K and $1.6$ times at 128K, and falls to $0.2$ times at 1M, where charging almost nothing beyond $10^{5}$ tokens leaves no margin against the one-fifth of rounds that miss. Cold events---a first prefill, the first request after an eviction and a provider transfer---fall outside the steady-state statement in any case.

\noindent Agent and routing workloads are set out in Appendix~D.1: agent traffic does not benefit automatically, and weighting is $96.8\%$ dearer once $5{,}000$ persistent tool tokens accumulate per round, which is the mechanism declining to discount observations that Section~\ref{sec:ctx} shows are irreplaceable, whereas a router that adopts the portable rule is never worse off---it pays the same when nothing moves and $18.5\%$ less when every round does. Robustness across the $180$-combination grid and $20$ seeds is in Appendix~D.2: weighting is cheaper in $33\%$ of combinations, and across seeds it lowers P99 and the coefficient of variation in every replication but the mean in every one. How compression interacts with the two tariffs is set out in Appendix~D.4---in short, compression lowers the cache charge by $17$--$34\%$ at 128K while raising the weighted charge by $12$--$14\%$, because a summary is billed as a fresh block at close to the full rate and resets the ages it replaces, and at 32K that makes weighting the dearer of the two.

\section{Discussion}
A positional discount is not an efficiency instrument, and the paper does not claim that it is: Proposition~4 shows that pricing each retained block at its marginal serving cost implements the efficient retention choice, whereas a weight based on age does not, because age is not the determinant of a block's contribution. The case for the positional instance rests on two other margins. The first is risk: for application developers the measured effect is a smaller upper tail, with P99 down $42.0\%$ and the coefficient of variation down $40.5\%$, which matters when the cost of a long session is convex in its length. The second is competition: a portable rule lowers a destination's cold-start charge and hence the friction binding a user to an incumbent's cache. Proposition~5 states the conditions under which those channels raise welfare. Neither margin is an efficiency claim, and neither is unconditional. A fuller treatment would model provider competition explicitly---how an incumbent responds when a rival adopts a portable tariff---which this paper does not attempt.

A portability rule must also be attractive to the provider that grants it, and adoption is not obviously self-interested, since a destination forgoes part of the cold-start charge it could otherwise collect. Two supports are available, and neither is modelled here: reciprocity, because every provider is both a destination and an origin, so a rule that lowers the cost of leaving raises switching and hence total demand; and an offsetting fee, since a destination could apply the weight to verifiable history while charging a one-off verification and migration fee of the kind already denoted $\kappa$, converting part of the forgone revenue into an explicit service charge. Adoption therefore remains open, and so does the deeper governance question: providers know more than customers about performance and cost, users can reorganise context strategically, and a learned weight could discriminate across content or customer groups.

A final qualification concerns aggregate use rather than allocation. The efficiency this paper measures is in serving: a cache hit cuts the compute cost of a repeated prefix by two orders of magnitude (Section~\ref{sec:cost}), and the tariff lowers the effective price of context further. If cheaper context induces more of it, aggregate compute and spending can rise even as unit prices fall---the rebound effect \citep{sorrell2008}, whose canonical case is the Jevons paradox \citep{jevons1865}. Whether the rebound dominates is not something this study measures, and the provider condition already sits in that regime, since it is satisfied only under a demand elasticity near $3$ (Section~\ref{sec:mechanism}). The same channel bears on coverage, since induced demand lengthens retained history and the coverage constraint bites hardest on the cold re-send.

\subsection*{Limitations}
Four qualifications bound the evidence. The serving measurements come from a small model on a consumer GPU, so they establish scaling relationships rather than the absolute cost of a frontier deployment, and the coverage ceiling inherits that proxy character. The ablation's primary sample is one model family, although a 50-contract replication reproduces both results; its code and agent components are also small, so the legal-contract result carries most of the weight. The switching experiment spans two providers, so the premium mixes a cache effect with tokeniser and list-price differences, reported separately rather than netted out. Finally, the portable-discount rule is a counterfactual no provider offers, and the simulation retains calibrated assumptions for demand response and the variable-cost share.

\section{Conclusion}
This paper reframes nonlinear long-context pricing as an empirical and mechanism-design problem. Correct accounting shows that the lifetime charge for one historical token and the historical component of a request may be bounded while total expenditure of a continuing session generally is not, and that sustainable pricing cannot be inferred from non-negative weights but needs evidence on serving cost and demand. The measurements and the calibrated simulation sharpen that picture: contribution is driven by content relevance rather than position, so the geometric weight is an auditable approximation rather than the utility gradient; a cache hit reduces serving cost by two orders of magnitude, which makes a discounted historical charge sustainable, while the robust gain across calibrated sessions is a smaller upper tail rather than a lower average bill; and the measured switching premium is real, with a portable discount recovering much of it while tokeniser and price differences must be reported rather than netted out.

\section*{Supplementary information}
Appendices A--F, Tables S1--S6 and Figure S1 are included below. The ancillary file \nolinkurl{replication_package.zip} contains the replication code, recorded data and experiment instructions.

\section*{Declarations}
\subsection*{Funding and acknowledgements}
This research was supported by the National Social Science Fund of China (General Program, 22BTJ021).
\subsection*{Author contributions}
Y.L. formulated the research questions, designed the methodology, and reviewed and revised the manuscript. Y.Y. conducted preliminary experiments and wrote the initial draft. Q.L. completed the subsequent research, including theoretical derivations, software implementation, experiments, data analysis, validation and visualization, and substantially revised the manuscript. H.W. provided funding support.
\subsection*{Competing interests}
The authors declare no competing interests.
\subsection*{Data availability}
The data supporting the reported measurements and simulations are provided in the ancillary file \nolinkurl{replication_package.zip}, including the result files underlying the tables and figures, raw cross-provider usage records, context-ablation prompts and evaluation outputs, and serving-benchmark logs. The legal-contract corpus is the publicly available CUAD dataset \citep{datacuad}; the provider price schedules are available from the sources cited in the manuscript \citep{datadeepseek,dataqwen}. The original third-party datasets remain subject to their respective access and reuse conditions.
\subsection*{Code availability}
The simulation, analysis and benchmark code is provided in the ancillary file \nolinkurl{replication_package.zip}, together with experiment-specific instructions and recorded results.

\clearpage
\section*{Tables}
\begin{table}[H]
\centering\small\setlength{\tabcolsep}{4pt}
\caption{The cost distribution of Table~\ref{tab:developer} under proactive compression ($\tau=0.90$, $\rho=0.10$) against hard truncation, as percentage changes relative to the cache tariff on the same $2{,}000$ users. Compression moves only sessions long enough to fill the window, so the median is untouched. \textbf{[calibrated]}}
\label{tab:compressiondist}
\begin{tabular}{@{}lrr@{}}
\toprule
Metric & Hard truncation & Proactive compression \\
\midrule
Mean & -11.4 & -4.6 \\
Median (P50) & +58.9 & +58.9 \\
P95 & -25.2 & -21.9 \\
P99 & -42.0 & \textbf{-32.8} \\
Coefficient of variation & -40.5 & -30.3 \\
\bottomrule
\end{tabular}
\end{table}
\clearpage
\begin{table}[H]
\centering\small\setlength{\tabcolsep}{4pt}
\caption{Simulated distribution of user cost ($N=2{,}000$, \texttt{qwen3.8-flash} schedule) \textbf{[calibrated]}}
\label{tab:developer}
\begin{tabular}{lrrr}
\toprule
Metric & Cache tariff & Context-weighted & Change (\%) \\
\midrule
Mean (USD) & 0.055 & 0.048 & -11.4 \\
Standard deviation & 0.154 & 0.081 & -47.3 \\
Median (P50) & 0.014 & 0.022 & \textbf{58.9} \\
P95 & 0.245 & 0.183 & -25.2 \\
P99 & 0.618 & 0.359 & \textbf{-42.0} \\
Coefficient of variation & 2.812 & 1.673 & -40.5 \\
\bottomrule
\end{tabular}
\end{table}
\clearpage
\begin{table}[H]
\centering\small
\caption{Calibrated results under the three official price schedules ($r=0.95$, $h=0.80$, 200 rounds). Every column except crossover is a change relative to the cache tariff on the same token stream. The 200-round column is the session-cost advantage; mean, P99 and CV are the 2{,}000-user Monte Carlo distribution; the last column is the share of the 180-combination grid on which the weighted tariff is cheaper. P99 and the coefficient of variation fall under every schedule; the mean falls only under \texttt{qwen3.8-flash} and rises under the two DeepSeek schedules. \textbf{[calibrated]}}
\label{tab:schedules}
\begin{tabular}{@{}lrrrrrr@{}}
\toprule
Schedule (cached/list) & Crossover & 200-round & Mean & P99 & CV & Grid \\
 & (rounds) & saving (\%) & (\%) & (\%) & (\%) & cheaper (\%) \\
\midrule
DeepSeek \textit{flash} (2\%) & 182 & 3.3 & \textbf{+10.3} & \textbf{-30.9} & -42.6 & 22 \\
DeepSeek \textit{v4-pro} (3.3\%) & 164 & 7.4 & +7.4 & -32.8 & -43.5 & 23 \\
Qwen \textit{3.8-flash} (12.5\%) & 111 & 27.8 & -11.4 & -42.0 & -40.5 & 33 \\
\bottomrule
\end{tabular}
\end{table}
\clearpage
\section*{Figures}
\clearpage
\clearpage
\clearpage
\appendix
\renewcommand{\tabularxcolumn}[1]{>{\raggedright\arraybackslash}p{#1}}
\renewcommand{\thetable}{S\arabic{table}}
\renewcommand{\thefigure}{S\arabic{figure}}
\renewcommand{\theequation}{S\arabic{equation}}
\setcounter{table}{0}
\setcounter{figure}{0}
\setcounter{equation}{0}
\setstretch{1.0}

\noindent This supplementary document contains six appendices: scenario-specific formulations (A), proofs of Propositions 1--5 (B), the simulation setup and price schedules (C), further simulation results (D), the stakeholder conditions (E) and extended related work (F). Section, table and figure numbers of the main text are cited as such; supplementary tables and figures retain their S prefixes.

\section{Scenario-Specific Formulations}
\label{app:scenarios}
The general tariff separates persistent state from ephemeral computation. This permits the same mechanism to represent conversational, agentic and routed multi-model systems without assigning one universal decay rate to every token type.

\subsection{Ordinary dialogue}
In a conventional dialogue, the persistent increment is $\Delta_k=I_k+O_k$. With constant $I$, $O$ and block size $\Delta$, let
\begin{equation}
J=\left\lfloor\frac{W-I}{\Delta}\right\rfloor
\end{equation}
be the number of complete historical blocks that fit in the window. Ignoring at most one partially retained boundary block, the steady historical charge is
\begin{equation}
\overline C_{\rm chat,hist}=P_h\Delta\frac{1-r^J}{1-r}.
\label{eq:chatsteady}
\end{equation}
The system therefore has two phases: a growth phase while the window fills and a stationary sliding-window phase after eviction begins. Total session expenditure remains asymptotically linear with slope equal to new-input, output and steady historical charges.

\subsection{Agentic workflows}
An agent produces visible messages $O_k$, persistent tool observations $T_k$, optional summaries $S_k$ and ephemeral reasoning tokens $R_k$. Its persistent increment is
\begin{equation}
\Delta_k^{\rm agent}=I_k+O_k+T_k+S_k,
\end{equation}
while $R_k$ is charged only in the current round. Allowing separate weights for dialogue, tools and summaries gives
\begin{align}
C_k^{\rm agent}={}&P_{\rm in}I_k+P_{\rm out}O_k+P_{\rm reason}R_k \\
&+P_h\sum_{j<k}\sum_{z\in\{d,t,s\}}A_{jk}^{z}
w_z(D_{jk},x_{jk};\theta_z).
\label{eq:agent}
\end{align}
Tool outputs may retain value longer than conversational text because later actions depend on earlier observations. Conversely, a verified summary can replace several old blocks and reset the physical context state, but its own creation and quality loss must be counted. Agent workloads therefore require modality-specific parameters and a quality constraint, not merely a larger token volume.

\subsection{Intelligent routing and frequent model switching}
Let $m_k\in\{1,\ldots,M\}$ denote the model selected at round $k$, possibly by a router. Provider/model $m$ has prices $P_{{\rm in},m}$, $P_{{\rm cache},m}$, output price $P_{{\rm out},m}$, context window $W_m$ and quality $q_m(x_k)$. Define the switch indicator $s_k=\ind(m_k\ne m_{k-1})$. A cache tariff depends on provider-local cache state $h_{mk}$:
\begin{equation}
C_k^{\rm route,cache}=P_{{\rm in},m_k}I_k+
\left[h_{m_k,k}P_{{\rm cache},m_k}+(1-h_{m_k,k})P_{{\rm in},m_k}\right]L_{m_k,k}
+P_{{\rm out},m_k}O_k.
\label{eq:routecache}
\end{equation}
After a switch, $h_{m_kk}$ may fall to zero even when the semantic history is unchanged. Under portable utility weighting,
\begin{equation}
C_k^{\rm route,w}=P_{{\rm in},m_k}I_k+
P_{h,m_k}\sum_{j<k}A_{jk}^{(m_k)}w(D_{jk},x_{jk};\theta)+P_{{\rm out},m_k}O_k+\kappa s_k,
\label{eq:routeweighted}
\end{equation}
where $\kappa$ is the verification, serialisation and migration cost. The router chooses a model to minimise expected quality-adjusted cost,
\begin{equation}
m_k^*=\arg\min_m\left\{C_k(m)+\lambda_q\,\ell(q_m,x_k)+\lambda_d\,d_m(x_k)\right\},
\label{eq:routerchoice}
\end{equation}
where $\ell$ is expected quality loss and $d_m$ latency. Cheap switching is valuable only when the gain in price or quality exceeds $\kappa$ and any cache loss. Frequent routing can otherwise create excessive migration overhead or oscillation between models.

\noindent The migration cost $\kappa$ is a design assumption rather than a measurement; Section~4 of the main text states its value and notes that this paper does not measure it. The cross-provider experiment of Section~5.3 of the main text measures the switching premium rather than $\kappa$.

\section{Proofs of Propositions 1--5}

\label{app:proofs}

\noindent\textbf{Evidence retained from the window.} The number of tokens of block $j$ still chargeable at round $k$ is the interval overlap
\begin{equation}
\begin{aligned}
A_{jk}&=\left[\min\{H_j,H_{k-1}\}-\max\{H_{j-1},b_k\}\right]_+,\\
b_k&=\max\{0,H_{k-1}-(W-I_k)\},\qquad [z]_+=\max\{z,0\},
\end{aligned}
\label{eq:overlap}
\end{equation}
so that $0\le A_{jk}\le\Delta_j$ and $\sum_{j<k}A_{jk}=\min\{H_{k-1},W-I_k\}$. When the dialogue exceeds the window, fully evicted blocks receive $A_{jk}=0$, the oldest boundary block is charged only for the fraction still present, and newer blocks remain eligible. This avoids the incorrect rule $\ind(H_j\le W)$, which can exclude new material merely because cumulative history is large.

\medskip
\noindent\textbf{Proposition 1.} Equation~\eqref{eq:overlap} implies $\sum_{j<k}A_{jk}\le W-I_k\le W$. Multiplying by $0\le w\le1$ gives a historical charge no greater than $P_hW$, and the remaining terms are bounded by assumption. Stationarity yields a time-invariant distribution of the retained state, whereas summing a positive stationary per-request cost produces linear rather than bounded cumulative growth. $\square$

\medskip
\noindent\textbf{Proposition 2.} Let $Z$ tokens that cannot all remain be compressed to $\rho Z$ summary tokens. Let the summary preserve expected discounted downstream value $\phi V_Z$ while eviction preserves zero, and let $F(\rho Z)$ be the discounted future charge and serving cost of retaining it. Relative to eviction, compression adds preserved value $\phi V_Z$ and adds the one-time compression cost plus the future summary cost, so it is preferred if and only if
\begin{equation}
\phi V_Z>K^{\rm comp}+F(\rho Z).
\label{eq:compressioncondition}
\end{equation}
If quality loss is written as $(1-\phi)V_Z$, the equivalent comparison includes the value lost under each policy. $\square$

\medskip
\noindent\textbf{Proposition 3.} Subtract the common fresh-input and output terms from the cold and portable charges and rearrange. $\square$

\medskip
\noindent\textbf{Proposition 4.} Because tariff payments are transfers among users, applications and providers, total surplus is
\begin{equation}
\mathcal W(x,m)=V_a(Q_m(x))-G_m(x)-S_a(m_{k-1},m),
\label{eq:welfare}
\end{equation}
where $V_a(Q_m(x))$ is the user's monetary value of quality, $G_m(x)$ the provider's variable serving cost and $S_a(m_{k-1},m)$ non-monetary switching friction. The efficient context choice satisfies, for each block $j$ retained at an interior solution,
\begin{equation}
V_a'(Q)\frac{\partial Q_m}{\partial x_j}
=\frac{\partial G_m}{\partial x_j}+\frac{\partial S_a}{\partial x_j}.
\label{eq:efficientcontext}
\end{equation}
The application maximises $V_a(Q_m(x))-B_m(x)-F_m$; setting $\partial B_m/\partial x_j=\partial G_m/\partial x_j+\partial S_a/\partial x_j$ makes its first-order condition identical to Equation~\eqref{eq:efficientcontext}. Since the fixed fee $F_m$ does not vary with $x$, it does not alter the marginal choice, so it can recover fixed cost without distorting block-level retention. $\square$

\medskip
\noindent\textbf{Proposition 5 (a decomposition).} Let regime $0$ be current cache pricing and regime $1$ the proposed mechanism, with provider profit $\Pi_z$, application unit margin $M_z$, user net utility $U_z$ and application cost-at-risk $\rho(C_z)$ for $z\in\{0,1\}$. The proposed regime is a weak Pareto improvement for the three focal stakeholders if
\begin{align}
&D_1(p_1-g_1)\ge D_0(p_0-g_0), &&\text{provider condition},\label{eq:paretoprovider}\\
&\mathbb E[M_1]\ge\mathbb E[M_0],\quad \rho(C_1)\le\rho(C_0), &&\text{application condition},\label{eq:paretoapp}\\
&U_1\ge U_0, &&\text{user condition},\label{eq:paretouser}
\end{align}
with at least one strict inequality. Writing $p_z$ and $g_z$ for expected revenue and serving cost per representative session and $D_z$ for demand, the provider condition is equivalently
\begin{equation}
D_1(p_1-g_1)>D_0(p_0-g_0).
\label{eq:providerbenefit}
\end{equation}

\medskip
\noindent\textbf{Proof.} Each of the three conditions is the statement that one stakeholder's payoff is weakly higher under regime $1$ than under regime $0$, evaluated on that stakeholder's own functional: the provider's $\Pi_z$; the application's $\mathbb E[M_z]$ and $\rho(C_z)$; and the user's $U_z$. A tariff is a transfer among these three parties and no fourth party is modelled, so the three inequalities together mean that no stakeholder is worse off, which is weak Pareto improvement; at least one strict inequality makes it strict for that stakeholder. Each condition is also necessary, since if it fails that stakeholder is strictly worse off. The provider condition is stated in demand form because $\Pi_z=D_z(p_z-g_z)$ by definition, which makes $\Pi_1\ge\Pi_0$ and $D_1(p_1-g_1)\ge D_0(p_0-g_0)$ the same inequality; the strict form is Equation~\eqref{eq:providerbenefit}. The proposition is therefore an equivalence rather than a substantive implication, and its content is the empirical question of whether the four inequalities hold. $\square$

\section{Simulation Setup and Price Schedules}

\label{app:implementation}

\subsection{Context compression as a dynamic state transition}
A finite window does not imply that an application waits until the window is full and then drops tokens mechanically. Long dialogues often compress proactively because attention dilution, latency and future storage charges can become material before the hard limit. Let $X_{k-1}$ denote tokens in the persistent state before round $k$, and let a memory controller choose an old-token volume $Z_k$ to summarise when $X_{k-1}>\tau(W-I_k-O_k)$, where $0<\tau\le1$ is a trigger and the bracketed term is the window less room for the current input and for the output that round will retain. A compression ratio $\rho\in(0,1)$ replaces $Z_k$ original tokens with $\rho Z_k$ summary tokens. Let $E_k$ be additional hard eviction. The post-round state obeys
\begin{equation}
X_k=X_{k-1}+I_k+O_k-(1-\rho)Z_k-E_k\le W.
\label{eq:statecompression}
\end{equation}
The compression call is not free. If it reads the selected history once and generates the summary, its one-time charge is
\begin{equation}
K_k^{\rm comp}=P_{\rm in}^{c}Z_k+P_{\rm out}^{c}\rho Z_k.
\label{eq:compressioncost}
\end{equation}
After replacement, removed originals have zero overlap in Equation~\eqref{eq:overlap}; the summary becomes a new persistent block and is charged on subsequent use. This convention prevents double billing. If $U(Z_k)$ is the task-relevant utility mass of the source and $\phi_k\in[0,1]$ is summary fidelity, the new block carries utility $\phi_kU(Z_k)$ rather than the full source utility. Thus compression changes both physical token state and information state.

The application chooses compression, eviction and retention dynamically:
\begin{equation}
\min_{\{Z_k,E_k\}}\;\mathbb E\sum_{k=1}^{N}\delta^{k-1}
\left[C_k^w+K_k^{\rm comp}+\lambda_q\mathcal L_k(\phi_k,Z_k,E_k)+\lambda_d d_k\right]
\quad\text{s.t. Equation~\eqref{eq:statecompression}},
\label{eq:memorycontrol}
\end{equation}
where $\mathcal L_k$ is downstream quality loss and $d_k$ is latency. Consequently, window size alone cannot determine the optimal policy: a larger $W$ delays forced eviction, but compression may still be optimal when its discounted future savings and preserved task value exceed its generation cost and fidelity loss.

The trigger $\tau$ is a property of the application rather than of the model, and deployed systems place it high. Most fire when the next call would overflow---at the window less a reserve for the response, which is $\tau\approx0.90$--$0.98$ for the window sizes studied here---and the remainder use a watermark between $0.75$ and $0.90$; a minority that also guard against oversized tool outputs use $0.50$. We therefore sweep $\tau\in[0.75,1.00]$ and report the range rather than a point. Three qualifications attach to that sweep. The trigger is a fraction of capacity here, whereas the largest family of deployed triggers is specified as a fixed reserve, which makes $\tau$ rise with $W$; holding $\tau$ fixed across the three windows is the fraction-type special case, and it is what the cross-window comparison assumes. Message-count and absolute-token triggers---compact after thirty messages, or at $150{,}000$ tokens---do not map onto a fraction of the window unless the per-round increment is constant, which this simulation assumes and a production agent violates whenever a tool result arrives in bulk. And because the simulation carries neither the quality term $\mathcal L_k$ nor the latency term $d_k$, it cannot identify an optimal $\tau$, and none is claimed.

Two further properties are visible in the artefact rather than in the text above, and Equation~\eqref{eq:memorycontrol} should be read accordingly: the run instantiates a fixed rule rather than solving that problem, so the equation is the control problem of which the rule is one crude instance, and its quality and latency terms are carried but never evaluated. A compression event removes the oldest \emph{half} of the retained history and replaces it with a summary $\rho$ times its length, so the working set falls to about $0.55$ of its previous size and the window is not reached again under this policy: compression displaces eviction rather than deferring it. And $\rho$ and $\phi$ are assumed independently---the model does not couple summary length to fidelity---so $\rho$ enters only the token counts and hence the charge, which makes the $\rho$ sweep one-directional, while $\phi$ enters only the retained-utility bookkeeping, which this paper does not report. Neither parameter identifies an interior optimum here, and none is claimed.

\subsection{Parameter grid and prices}
\begin{table}[H]
\centering\small
\caption{Parameters of the calibrated simulation experiments}
\label{tab:parameters}
\begin{tabularx}{\textwidth}{@{}llX@{}}
\toprule
Component & Baseline value & Robustness values or distribution \\
\midrule
Interaction rounds & 200 & Fixed at 200 in the grid \\
New input per round & 250 tokens & 50, 581, 1,000 or 2,000 \\
Retained output per round & 750 tokens & 1,120, 1,200, 1,500 or 5,000 \\
Context window $W$ & 128,000 tokens & 32,000; 128,000 \\
Cache-hit rate $h$ & 0.80 & 0.50; 0.80; 0.95 \\
Utility parameter $r$ & 0.95 & 0.70; 0.80; 0.90; 0.95; 0.98; 0.99 \\
Uncached input price & 0.8 CNY/million & Fixed in calibrated analysis \\
Cached input price & 0.10 CNY/million & Fixed in calibrated analysis \\
Output price & 2.7 CNY/million & Fixed in calibrated analysis \\
Historical base price $P_h$ & Equal to the uncached input price & Fixed in calibrated analysis \\
Baseline provider sessions & 10,000 & Fixed in sensitivity analysis \\
Provider variable cost & 55\% of baseline revenue & Assumed, not estimated \\
Demand elasticity & 0--3 in increments of 0.1 & 31 sensitivity points \\
Monte Carlo users & 2,000 & 20 replications of 500 users \\
Primary random seed & 20,260,913 & Seeds 20,260,901--20,260,920 \\
Memory-policy rounds & 1,200 & Fixed so the 1M window is reached; 9{,}600 for the steady-state check \\
Compression trigger $\tau$ & 0.90 & 0.75; 0.80; 0.90; 1.00 \\
Compression ratio $\rho$ & 0.10 & 0.05; 0.10; 0.20 \\
Summary fidelity $\phi$ & 0.85 & 0.70; 0.85; 0.95 \\
\bottomrule
\end{tabularx}
\end{table}

\begin{table}[H]
\centering\footnotesize
\caption{Official price schedules used in the calibrated analysis}
\label{tab:pricesource}
\begin{tabularx}{\textwidth}{@{}p{.20\textwidth}rrrX@{}}
\toprule
Category & DeepSeek & DeepSeek & Qwen & \\
 & \textit{flash} (idle/peak) & \textit{v4-pro} (idle/peak) & \textit{3.8-flash} & \\
\midrule
Uncached input & 1 / 2 & 4.5 / 9.0 & 0.8 & CNY / million tokens \\
Cached input & 0.02 / 0.04 & 0.15 / 0.30 & 0.1 & CNY / million tokens \\
Output & 4 / 8 & 13.5 / 27.0 & 2.7 & CNY / million tokens \\
\bottomrule
\end{tabularx}
\end{table}

\noindent\textit{Sources.} DeepSeek, ``Models \& Pricing,'' \url{https://api-docs.deepseek.com/zh-cn/quick_start/pricing/}, models \texttt{deepseek-flash} (V4.1-Flash) and \texttt{deepseek-v4-pro}, context 1M, accessed 2026-09-13; idle applies outside weekday 09:00--12:00 and 14:00--18:00 Beijing and is half of peak. Alibaba Cloud DashScope, ``Model pricing,'' \url{https://help.aliyun.com/zh/model-studio/model-pricing}, model \texttt{qwen3.8-flash}, context 1M, accessed 2026-09-13; the input price was reduced from 1.0 to 0.8 CNY on 2026-08-27. Convert CNY to USD at the rate reported in the data-availability package. The two vendors are the independent providers used for the cross-provider experiment (Section~5.3 of the main text).

\section{Further Simulation Results}

\subsection{Agents and intelligent routing}
Additional simulations separate persistent tool observations from ephemeral reasoning. Agent workloads do not benefit automatically---the weighted tariff is $8.6\%$ more expensive than cached pricing for a 100-round agent with no tool tokens, and $96.8\%$ more expensive once $5{,}000$ persistent tool tokens accumulate per round---but on this evidence that is the mechanism declining to discount rather than a defect: Section~5.1 of the main text finds that every early observation carries a contribution near one and no age signal, so a tariff that discounted persistent tool state by age would be discounting precisely what should not be discounted. The loss appears because tool output is charged at the uncached input price while the benchmark reuses a strong cache, which is a statement about the historical base price for tool tokens rather than about the weighting. Intelligent routing does not have that problem: a router that adopts the portable rule is never worse off than one that relies on the incumbent's cache, paying the same when nothing is transferred and $18.5\%$ less when every round is, because the rule prices only history that actually moves. Table~S3 gives the full grid, with changes again measured against the cache tariff on the same token stream.

Every row shares the same 100-round token stream except for the parameter named in the middle column, and each change is measured against the cache tariff on that stream.
\begin{table}[H]
\centering\small
\caption{Selected calibrated results for agents and intelligent routing}
\label{tab:agentrouter}
{\setlength{\tabcolsep}{4.5pt}\begin{tabular}{llrrr}
\toprule
Scenario & Parameter & Cache cost & Weighted cost & Change \\
\midrule
Agent, 100 rounds & No tool/reasoning tokens & 0.289 & 0.314 & 8.6\% \\
Agent, 100 rounds & 5,000 persistent tool tokens & 0.473 & 0.931 & \textbf{96.8\%} \\
Agent, 100 rounds & +5,000 ephemeral reasoning tokens & 0.663 & 1.121 & 69.1\% \\
Router, per round & Switch probability 0.75 & 0.00905 & 0.00749 & -17.3\% \\
Router, per round & Switch probability 1.00 & 0.01127 & 0.00918 & \textbf{-18.5\%} \\
\bottomrule
\end{tabular}}
\end{table}

\subsection{Workload, window, cache-rate and discount grid}
Robustness is mixed: on the \texttt{qwen3.8-flash} schedule, utility weighting was cheaper in 60 of the 180 grid combinations (33\%), with a median expenditure change of $+36.3\%$; across 20 seeds it reduced P99 and the coefficient of variation in every replication, but not the mean in every replication. Table~S4 gives the full grid and the seed replications. The simulation's own transfer premium for a $100{,}000$-token history is $370.6\%$, below the measured figure because it prices both providers on one schedule.

The grid crosses five workload types, two windows, three cache-hit rates and six values of $r$; the seed rows repeat a 500-user Monte Carlo sample over 20 fixed seeds.
\begin{table}[H]
\centering\small
\caption{Robustness results relative to the cache tariff (the grid row uses the \texttt{qwen3.8-flash} schedule; Table~3 of the main text reports the other two)}
\label{tab:robustness}
\begin{tabular}{lrrr}
\toprule
Outcome & Mean/median & Minimum & Maximum \\
\midrule
Grid expenditure change (180 combinations) & \textbf{36.3\%}$^{a}$ & -84.1\% & 300.3\% \\
Seed mean-cost change (20 replications) & -8.4\% & -18.8\% & 0.2\% \\
Seed P95 change (20 replications) & -22.0\% & -33.7\% & 1.5\% \\
Seed P99 change (20 replications) & \textbf{-40.6\%} & -53.8\% & -19.0\% \\
Seed coefficient-of-variation change & -32.5\% & -46.4\% & -16.4\% \\
\bottomrule
\multicolumn{4}{l}{\footnotesize $^{a}$Median; seed rows report arithmetic means. Negative values indicate savings or risk reduction.} \\
\end{tabular}
\end{table}

\subsection{Proactive compression over the trigger range}
The trigger $\tau$ is an application parameter rather than a model parameter, and Table~\ref{tab:compressiongrid} reports the policy across the range deployed systems use, at both the $1{,}200$-round horizon of Section~5.4 of the main text and a $9{,}600$-round horizon long enough for the working set to complete a compression cycle at every window.

\begin{table}[H]
\centering\small
\caption{Proactive compression across the trigger range, as a percentage change in the cumulative charge under each tariff relative to hard truncation on the same token stream ($\rho=0.10$, $\phi=0.85$). The 1M entries of the left block are transients: the window is first reached at round $1{,}001$ of $1{,}200$, so the working set has not completed a compression cycle. \textbf{[calibrated]}}
\label{tab:compressiongrid}
\begin{tabular}{@{}llrrrr@{}}
\toprule
& & \multicolumn{2}{c}{1{,}200 rounds} & \multicolumn{2}{c}{9{,}600 rounds} \\
\cmidrule(lr){3-4}\cmidrule(lr){5-6}
Window & $\tau$ & Cache & Weighted & Cache & Weighted \\
\midrule
32K & 0.75 & -19.8 & -10.1 & -20.1 & -10.3 \\
32K & 0.80 & -17.2 & -7.7 & -17.4 & -7.9 \\
32K & 0.90 & -10.1 & -1.8 & -10.3 & -1.9 \\
32K & 1.00 & -5.9 & +1.4 & -6.0 & +1.3 \\
\midrule
128K & 0.75 & -34.2 & +12.2 & -35.4 & +12.9 \\
128K & 0.80 & -30.5 & +12.8 & -31.6 & +13.6 \\
128K & 0.90 & -23.4 & +13.6 & -24.5 & +14.7 \\
128K & 1.00 & -16.6 & +14.3 & -17.3 & +15.4 \\
\midrule
1M & 0.75 & -23.4 & +9.7 & \textbf{-39.6} & +16.2 \\
1M & 0.80 & -19.0 & +10.0 & -35.9 & +16.0 \\
1M & 0.90 & -14.0 & +5.8 & -28.3 & +15.8 \\
1M & 1.00 & -9.5 & +6.5 & \textbf{-20.7} & +16.0 \\
\bottomrule
\end{tabular}
\end{table}

\subsection{Truncation, compression and the charge}
\label{sec:compression}
Every scenario above reaches the window limit: at the baseline volume, the 128K window is full from round 129 of 200, after which the per-round charge stops rising and the curves of Figure~2(a) of the main text flatten. What happens next depends on how the application disposes of history it can no longer hold. Eviction is not charged at all---a token that leaves the window is never billed again, so the request charge is bounded by $P_{\rm in}I+P_hW+P_{\rm out}O$ (Proposition~1)---whereas compression is paid for once, as an input pass over the compressed tokens plus the summary tokens it generates, after which the summary is billed as an ordinary persistent block. Compression is proactive rather than a response to overflow: the controller summarises once occupancy passes a fraction $\tau$ of the window, which is why the dash-dot line in the figure precedes the dotted one (Appendix~C).

Figure~S1 follows a 1{,}200-round session under both policies and both charging methods, and Table~S5 gives the trigger grid. Until compression first triggers---round 116 at 128K and round 901 at 1M, both before the window binds at 129 and 1{,}001---the four curves coincide; afterwards they separate, and the separation runs in opposite directions for the two tariffs. Cached summaries cut the cache charge at every trigger an application is likely to use ($\tau\in[0.75,1.00]$): by $6$--$20\%$ at 32K, $17$--$34\%$ at 128K and $10$--$23\%$ at 1M, because compression lowers the retained working set and the cache tariff re-bills that working set every round. Over the 1{,}200 rounds at 128K that is a fall from 5.253 to 4.025. The weighted charge moves the other way, because a summary is billed as a fresh block near the full rate and resets the ages it replaces: it rises from 2.899 to 3.293 at 128K, that is $12$--$14\%$ over the range, and $6$--$10\%$ at 1M, while at 32K it changes sign inside the band. Those rises do not make the weighted tariff the dearer of the two---at 128K and 1M its absolute charge stays well below the cache tariff (\$3.29 against \$4.03; \$3.07 against \$20.7)---but at 32K it is above it (\$2.36 against \$1.45). The compression calls themselves are a minor item, about $4\%$ of the session total at 128K. The 1M entries of a 1{,}200-round run are transients: the window is first reached at round $1{,}001$, and over $9{,}600$ rounds the saving roughly doubles. The mechanism therefore buys its advantage where history is long and does not need to be compressed, and pays for it as the more expensive of the two tariffs when the window is short.

\begin{figure}[htp]
\centering
\includegraphics[width=\textwidth]{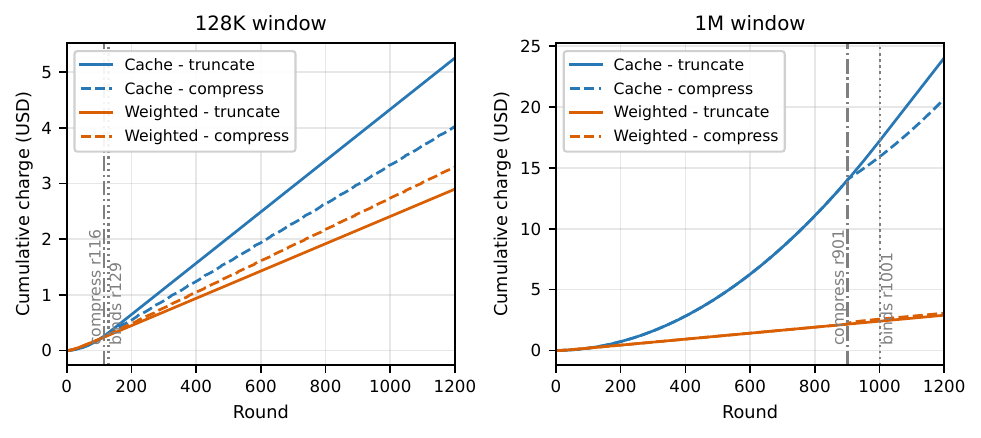}
\caption{Cumulative charge along a 1{,}200-round session, for the cache and the weighted tariff under hard truncation (solid) and under compression (dashed), at a 128K and a 1M window. Two vertical lines mark the two events that matter: the dash-dot line is the round at which compression first triggers (occupancy passing $\tau$ of capacity), and the dotted line the round at which the window first overflows; compression therefore displaces truncation rather than deferring it. Compression triggers at $\tau=0.90$ of capacity, at ratio $\rho=0.10$ and fidelity $\phi=0.85$; 250 new input and 750 retained output tokens per round, $h=0.80$, $r=0.95$. These are calibrated runs, not measurements.}
\label{fig:compression-path}
\end{figure}

\label{app:secondary}

\section{Stakeholder conditions and the measures that assess them}

Table~S6 restates the three conditions of Section~4 of the main text and the measure on which each outcome is assessed.

\begin{table}[H]
\centering\small
\caption{Stakeholder conditions and the measures that assess them (restates Section~4 of the main text)}
\label{tab:benefitsSI}
\begin{tabularx}{\textwidth}{@{}>{\raggedright\arraybackslash}p{.17\textwidth}>{\raggedright\arraybackslash}p{.30\textwidth}X@{}}
\toprule
Stakeholder & Required condition & Main outcome \\
\midrule
Model provider & Demand growth exceeds the usage-growth threshold of Section~4 of the main text and cost remains covered; the calibrated analysis requires \textbf{a demand elasticity near $3$}, atypically high for a mature digital service & Sessions, rounds, retention, contribution profit \\
Application developer & Per-round historical charge stabilises and quality is preserved & Mean, CV, P95/P99, CVaR, task quality \\
End user & Portable and auditable context-age/utility metadata & Switching premium, completion rate, provider switching \\
\bottomrule
\end{tabularx}
\end{table}

\section{Extended Related Work}
\label{app:related}
\subsection{Digital-service pricing}
Cloud and software markets have moved toward subscriptions and usage-based tariffs \citep{cusumano2010}. Work on digital goods emphasises nonlinear pricing, demand heterogeneity and the gap between measured productivity and user surplus \citep{brynjolfsson2017,frederick2002}.

Nonlinear and two-part tariffs are the classical instruments for pricing a divisible good when marginal valuation varies across units or buyers. The theory begins with Mussa and Rosen \citep{mussa1978}, Maskin and Riley \citep{maskin1984} and Wilson's synthesis \citep{wilson1993}; Oi \citep{oi1971} introduced the two-part tariff and Rochet and Stole \citep{rochet2002} extended nonlinear pricing to random participation, while Armstrong \citep{armstrong1999} and Rochet and Tirole \citep{rochet2006} connect these instruments to multi-product and two-sided settings.

Information goods require a distinct treatment because marginal reproduction cost is near zero while value is heterogeneous \citep{shapiro1999,bakos1999,sundararajan2004}, and for cloud and software markets Choudhary \citep{choudhary2007} compares perpetual licensing with subscription or usage-based schemes while Armbrust et al. \citep{armbrust2010}, Buyya et al. \citep{buyya2009} and Weinhardt et al. \citep{weinhardt2009} describe the technical and business architecture such tariffs must respect. Platform governance and control of an intermediated service are examined by Tiwana et al. \citep{tiwana2010}, Boudreau \citep{boudreau2010}, Hagiu and Wright \citep{hagiu2015} and Parker et al. \citep{parker2016}; more recently, Bergemann, Bonatti and Smolin \citep{bergemann2025} study token allocation, fine-tuning and optimal LLM pricing directly.


\clearpage
\begin{thebibliography}{99}
\bibitem{vaswani2017} Vaswani, A., Shazeer, N., Parmar, N., et al.: Attention is all you need. \emph{Advances in Neural Information Processing Systems}, 30 (2017).

\bibitem{dao2023} Dao, T.: FlashAttention-2: Faster attention with better parallelism and work partitioning. arXiv:2307.08691 (2023).

\bibitem{kwon2023} Kwon, W., Li, Z., Zhuang, S., et al.: Efficient memory management for large language model serving with PagedAttention. In \emph{Proceedings of SOSP 2023} (2023).

\bibitem{child2019} Child, R., Gray, S., Radford, A., and Sutskever, I.: Generating long sequences with sparse transformers. arXiv:1904.10509 (2019).

\bibitem{acharya2025} Acharya, S., Jia, F., and Ginsburg, B.: Star Attention: Efficient LLM inference over long sequences. In \emph{Proceedings of the 42nd International Conference on Machine Learning (ICML)}, PMLR 267, 356--371 (2025).

\bibitem{lin2025} Lin, Z., Nikishin, E., He, X. O., and Courville, A.: Forgetting Transformer: Softmax attention with a forget gate. arXiv:2503.02130 (2025).

\bibitem{wang2024} Wang, L., Ma, C., Feng, X., et al.: A survey on large language model based autonomous agents. \emph{Frontiers of Computer Science}, 18, 186345 (2024).

\bibitem{xi2025} Xi, Z., Chen, W., Guo, X., et al.: The rise and potential of large language model based agents: A survey. \emph{Science China Information Sciences}, 68, 121101 (2025).

\bibitem{boiko2023} Boiko, D. A., MacKnight, R., Kline, B., et al.: Autonomous chemical research with large language models. \emph{Nature}, 624, 570--578 (2023).

\bibitem{li2023} Li, Y., Wang, S., Ding, H., and Chen, H.: Large language models in finance: A survey. In \emph{Proceedings of ICAIF 2023}, 374--382 (2023).

\bibitem{thirunavukarasu2023} Thirunavukarasu, A. J., Ting, D. S. J., Elangovan, K., et al.: Large language models in medicine. \emph{Nature Medicine}, 29, 1930--1940 (2023).

\bibitem{kasneci2023} Kasneci, E., Sessler, K., Kuechemann, S., et al.: ChatGPT for good? On opportunities and challenges of large language models for education. \emph{Learning and Individual Differences}, 103, 102274 (2023).

\bibitem{kazemitabaar2023} Kazemitabaar, M., Chow, J., Ma, C. K. T., et al.: Studying the effect of AI code generators on supporting novice learners in introductory programming. In \emph{Proceedings of the 2023 CHI Conference on Human Factors in Computing Systems}, Article 455, 1--23 (2023).

\bibitem{katz2024} Katz, D. M., Bommarito, M. J., Gao, S., and Arredondo, P.: GPT-4 passes the bar exam. \emph{Philosophical Transactions of the Royal Society A}, 382, 20230254 (2024).

\bibitem{bergemann2025} Bergemann, D., Bonatti, A., and Smolin, A.: The economics of large language models: Token allocation, fine-tuning, and optimal pricing. arXiv:2502.07736 (2025).

\bibitem{mussa1978} Mussa, M., and Rosen, S.: Monopoly and product quality. \emph{Journal of Economic Theory}, 18(2), 301--317 (1978).

\bibitem{maskin1984} Maskin, E., and Riley, J.: Monopoly with incomplete information. \emph{RAND Journal of Economics}, 15(2), 171--196 (1984).

\bibitem{wilson1993} Wilson, R. B.: \emph{Nonlinear Pricing}. Oxford University Press (1993).

\bibitem{oi1971} Oi, W. Y.: A Disneyland dilemma: Two-part tariffs for a Mickey Mouse monopoly. \emph{Quarterly Journal of Economics}, 85(1), 77--96 (1971).

\bibitem{rochet2002} Rochet, J.-C., and Stole, L. A.: Nonlinear pricing with random participation. \emph{Review of Economic Studies}, 69(1), 277--311 (2002).

\bibitem{armstrong1999} Armstrong, M.: Price discrimination by a many-product firm. \emph{Review of Economic Studies}, 66(1), 151--168 (1999).

\bibitem{rochet2006} Rochet, J.-C., and Tirole, J.: Two-sided markets: A progress report. \emph{RAND Journal of Economics}, 37(3), 645--667 (2006).

\bibitem{shapiro1999} Shapiro, C., and Varian, H. R.: \emph{Information Rules: A Strategic Guide to the Network Economy}. Harvard Business School Press (1999).

\bibitem{bakos1999} Bakos, Y., and Brynjolfsson, E.: Bundling information goods: Pricing, profits, and efficiency. \emph{Management Science}, 45(12), 1613--1630 (1999).

\bibitem{sundararajan2004} Sundararajan, A.: Managing digital piracy: Pricing and protection. \emph{Information Systems Research}, 15(3), 287--308 (2004).

\bibitem{choudhary2007} Choudhary, V.: Comparison of software quality under perpetual licensing and software as a service. \emph{Journal of Management Information Systems}, 24(2), 141--165 (2007).

\bibitem{armbrust2010} Armbrust, M., Fox, A., Griffith, R., et al.: A view of cloud computing. \emph{Communications of the ACM} 53(4), 50--58 (2010).

\bibitem{buyya2009} Buyya, R., Yeo, C. S., Venugopal, S., Broberg, J., and Brandic, I.: Cloud computing and emerging IT platforms. \emph{Future Generation Computer Systems}, 25(6), 599--616 (2009).

\bibitem{weinhardt2009} Weinhardt, C., Anandasivam, A., Blau, B., et al.: Cloud computing---a classification, business models, and research directions. \emph{Business \& Information Systems Engineering}, 1(5), 391--399 (2009).

\bibitem{tiwana2010} Tiwana, A., Konsynski, B., and Bush, A. A.: Platform evolution: Coevolution of platform architecture, governance, and environmental dynamics. \emph{Information Systems Research}, 21(4), 675--687 (2010).

\bibitem{boudreau2010} Boudreau, K. J.: Open platform strategies and innovation: Granting access vs. devolving control. \emph{Management Science}, 56(10), 1849--1872 (2010).

\bibitem{hagiu2015} Hagiu, A., and Wright, J.: Multi-sided platforms. \emph{International Journal of Industrial Organization}, 43, 162--174 (2015).

\bibitem{parker2016} Parker, G. G., Van Alstyne, M. W., and Choudary, S. P.: \emph{Platform Revolution}. W. W. Norton (2016).

\bibitem{liu2024} Liu, N. F., Lin, K., Hewitt, J., Paranjape, A., Bevilacqua, M., Petroni, F., and Liang, P.: Lost in the middle: How language models use long contexts. \emph{Transactions of the Association for Computational Linguistics}, 12, 157--173 (2024).

\bibitem{sorrell2008} Sorrell, S., and Dimitropoulos, J.: The rebound effect: Microeconomic definitions, limitations and extensions. \emph{Ecological Economics}, 65(3), 636--649 (2008).

\bibitem{jevons1865} Jevons, W. S.: \emph{The Coal Question: An Inquiry Concerning the Progress of the Nation, and the Probable Exhaustion of Our Coal-Mines}. Macmillan (1865).

\bibitem{datacuad} Hendrycks, D., Burns, C., Chen, A., \& Ball, S.: \emph{CUAD: An expert-annotated NLP dataset for legal contract review} [Data set]. \url{https://github.com/TheAtticusProject/cuad} (2021).

\bibitem{datadeepseek} DeepSeek.: \emph{Models \& pricing} [Data set]. \url{https://api-docs.deepseek.com/zh-cn/quick_start/pricing/} (2026).

\bibitem{dataqwen} Alibaba Cloud.: \emph{Model pricing} [Data set]. DashScope. \url{https://help.aliyun.com/zh/model-studio/model-pricing} (2026).

\bibitem{cusumano2010} Cusumano, M. A.: Cloud computing and SaaS as new computing platforms. \emph{Communications of the ACM} 53(4), 27--29 (2010).

\bibitem{brynjolfsson2017} Brynjolfsson, E., Rock, D., and Syverson, C.: Artificial intelligence and the modern productivity paradox. NBER Working Paper 24001 (2017).

\bibitem{frederick2002} Frederick, S., Loewenstein, G., and O'Donoghue, T.: Time discounting and time preference: A critical review. \emph{Journal of Economic Literature}, 40(2), 351--401 (2002).
\end{thebibliography}
\end{document}